\documentclass{svmult}

\usepackage{makeidx}         
\usepackage{graphicx}        
\usepackage{multicol}        

\makeindex             

\begin{document}

\title*{Topological defects and electronic
             properties in graphene.}

\author{Alberto Cortijo,\inst{1}\and
M.A.H. Vozmediano\inst{2}}
\institute{Instituto de  Ciencia de Materiales de Madrid, CSIC,
 Cantoblanco E28049 Madrid, Spain
\texttt{cortijo@icmm.csic.es} \and Grupo de Modelizaci\'on y
Simulaci\'on Num\'ericas, Universidad Carlos III de
Madrid, E28913, Legan\'es, Madrid, Spain
\texttt{vozmediano@icmm.csic.es}}

\maketitle

In this work we will focus on the effects produced by topological
disorder  on the electronic properties
of a graphene plane. The presence of this type of disorder induces
curvature in the samples of this material, making
quite difficult the application of standard techniques of many body
quantum theory. Once we understand the nature of these defects, we
can apply ideas belonging to quantum field theory in curved
space-time and extract  information on
physical properties that can be
measured experimentally.

\section{Introduction}
\label{intro}

Graphene is a two dimensional material formed by isolated layers of
carbon atoms arranged in a honeycomb-like lattice.

Each carbon atom is linked to three nearest neighbors due to the
$sp^{2}$ hybridization process, which leads to three strong $\sigma$
bonds in a plane and a partially filled $\pi$ bond, perpendicular to the plane.
These $\pi$ bonds will determine  the low energy
electronic and transport properties of the system.
\\
\\
It is possible to derive a long wavelength tight binding hamiltonian
for the electrons in these $\pi$ bonds(\cite{W47}). This hamiltonian
is:

\begin{equation}
H=-iv_{F}\int d^{2}\textbf{r}\bar{\Psi}(\textbf{r})\gamma^{j}
\partial_{j}\Psi(\textbf{r}),\label{hamiltonian}
\end{equation}
where $v_{F}$ being a constant with dimensions of velocity
($v_{F}\sim10^{3} m/s$). The wave equation derived from the
hamiltonian (\ref{hamiltonian}) is the Dirac equation in two
dimensions with the coefficients $\gamma^{j}$ being an
appropriate set of
Dirac matrices. We can set for instance, $\gamma^{1}=\mathbf{1} \otimes\sigma_{1}$ and
$\gamma^{2}=\tau_{3}\otimes\sigma_{2}$,
where the $\sigma, \tau$ matrices are related to the
sublattice and Fermi point degrees of freedom respectively. The unexpected form of
the tight-binding Hamiltonian comes from two special features of the
honeycomb lattice:
first, the unit cell contains two carbon atoms belonging to
different triangular sublattices, and second, in the neutral
system at half filling, the Fermi surface reduces to two nonequivalent Fermi
points.  We will study the low energy states around any of these
two Fermi points. The dispersion relation obtained from
(\ref{hamiltonian}) is $\varepsilon(\textbf{k})=\pm
v_{F}|\textbf{k}|$, leading to a constant density of states,
$\rho^{0}(\omega)=\frac{8}{\pi}|\omega|$.

\section{A first model for the topological defects in graphene}
\label{defects}

Several types of defects like vacancies, adatoms, complex
boundaries, and structural or topological defects have been observed
experimentally in the graphene lattice(\cite{Hetal04}) and studied theoretically (see for
example \cite{Leh1},\cite{VLSG05}, \cite{LSGV06}).
\\
Topological defects are produced by substitution of an
hexagonal ring of the honeycomb lattice by an n-sided polygon
with any n. Their presence impose non-trivial
boundary conditions on the electron wave functions which are difficult
to handle. A proposal made in \cite{GGV92} was to trade the boundary conditions
imposed by pentagonal defects by the presence of appropriate gauge fields coupled to the spinor
wave function. A generalization of this approach
to include various topological defects was presented in
\cite{LC04}. The strategy consists of determining the phase of the
gauge field by parallel transporting the
state in suitable form along a closed curve surrounding all the
defects.
\begin{equation}
\Psi(\theta=0)=T_{C}\Psi(\theta=2\pi)\Leftrightarrow\Psi(\theta=0)=
\exp({\oint_{_{C}}\textbf{A}_{a}T^{a}d\textbf{r}})\Psi(\theta=2\pi),\label{boundary}
\end{equation}
where $\textbf{A}_{a}$ are a set of gauge fields and $T^{a}$ a set
of matrices related to the pseudospin degrees of freedom of the
system.
\\
When dealing with multiple defects, we must consider a curve
surrounding all of them, as the one sketched in figure (1):
\begin{figure}[h]
  \begin{center}
\includegraphics[height=5.5cm]{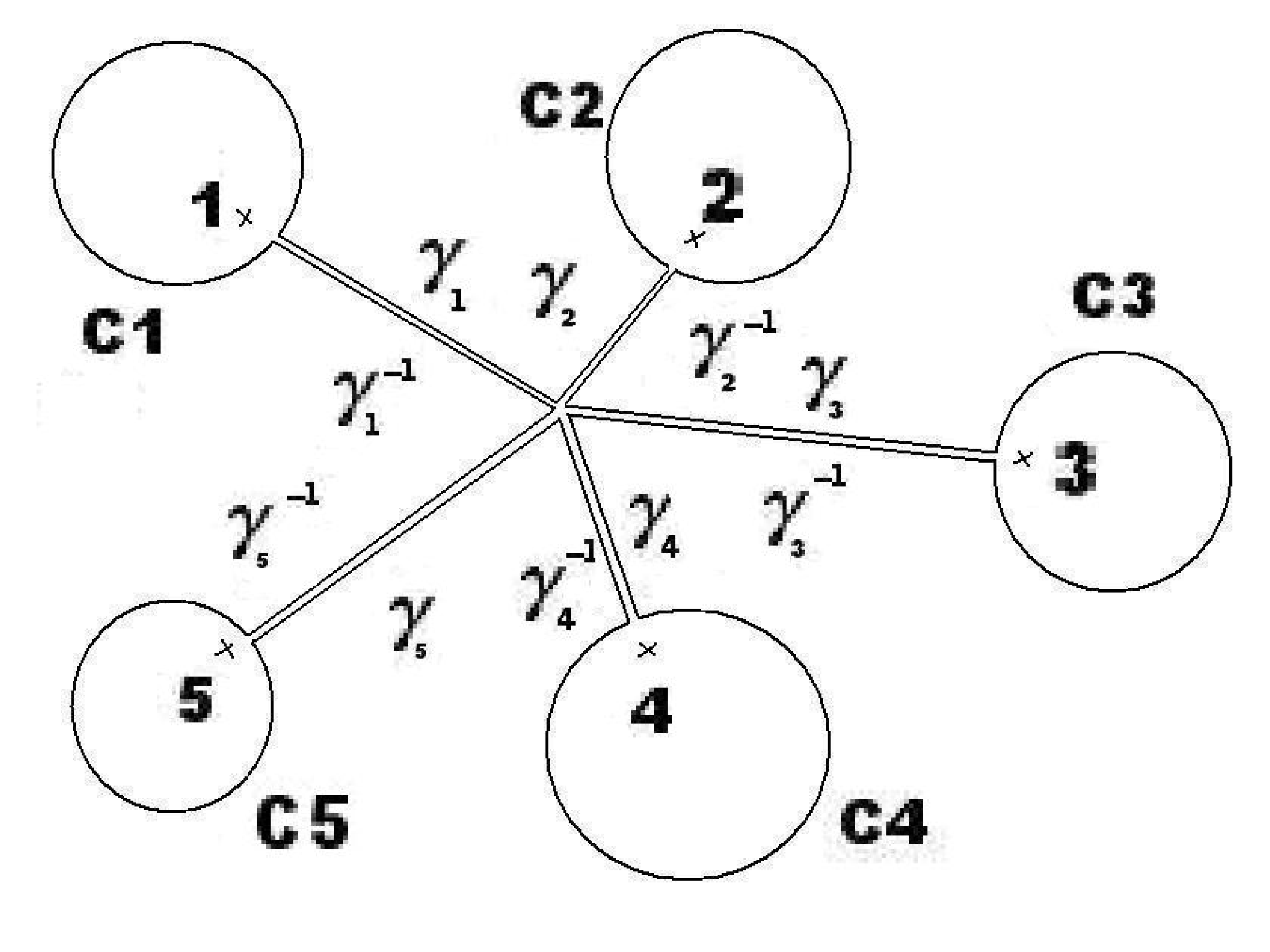}\label{escheme}
\end{center}
\caption{Scheme of a prototypical curve enclosing multiple defects
in which the state will be parallel transported.}
\end{figure}

The contour C is made of closed circles enclosing each defect
and straight paths linking all the contours to a fixed origin.
The parallel transport operator $P_{C}$ associated to the closed
path is thus a composition of transport operators over each piece:

\begin{equation}
P=P_{\gamma1}\cdot P_{1}\cdot P^{-1}_{\gamma1}\cdot...\cdot
P_{\gamma N}\cdot P_{N}\cdot P^{-1}_{\gamma N}.\label{composite}
\end{equation}

As explained in \cite{LC04}the total holonomy turns out to
be\footnote{The usual chiral lattice real vector basis for the
honeycomb lattice is used in this derivation.}:

\begin{equation}
P=(i)^{N}(\tau_{2})^{N}exp\left(\frac{2\pi
i}{6}(N^{+}-N^{-})\sigma_{3}\right) exp\left(\frac{2\pi
i}{3}\sum^{N}_{j=1}(n_{j}-m_{j})\tau_{3}\right)
 .\label{composite2}
\end{equation}

From equation (\ref{composite2}) we see that we have in principle
three different gauge fields to incorporate into the Dirac equation, which
couple to the matrices $\sigma_{3}$, $\tau_{2}$, and $\tau_{3}$ and
whose associated fluxes are adjusted from (\ref{boundary}).

\section{Generalization of the model}
\label{curvedmodel}

In spite of its elegance, the model presented in the previous
section does not contain the effects due to the curvature of
the layer in the presence of these defects. The model can be generalized
to account for curvature effects
(\cite{GGV92}, \cite{KO99}) by coupling  the gauge theory
obtained from the analysis of the holonomy in a curved
space.
\\
\\
The substitution of a hexagon by a polygon with $n<6$ sides gives
rise to a conical singularity with deficit angle $(2\pi/6)(6-n)$,
which is similar to the singularity generated by a cosmic string in
general relativity. The Dirac Equation for a massless spinor in a
curved spacetime is (\cite{birrell}):

\begin{equation}
i\gamma^{\mu}(x)(\partial_{\mu}-\Gamma^{(T)}_{j\mu})\psi=0,\label{curveddirac}
\end{equation}
where $\Gamma^{(T)}_{j\mu}$ is a set of spin connections related to
the pseudospin matrices in (\ref{composite2}) and $\gamma^{\mu}(x)$
are generalized Dirac matrices satisfying the anticommutation
relations
\begin{equation}
\left\{\gamma^{\mu}(x),\gamma^{\nu}(x)\right\}=2g^{\mu
\nu}(x).\label{commutation}
\end{equation}

The metric tensor in (\ref{commutation}) corresponds to a curved
spacetime generated by an arbitrary number of $N$ parallel cosmic
strings placed in $(a_{i},b_{i})$ (here we will follow the formalism
developed in \cite{AHO97}):

\begin{equation}
ds^{2}=-dt^{2}+e^{-\Lambda(x,y)}(dx^{2}+dy^{2}),\label{arcelement}
\end{equation}
with
$\Lambda(x,y)=\sum^{N}_{i=1}4\mu_{i}\log([(x-a_{i})^{2}+(x-b_{i})^{2}]^{1/2})$.
The parameters $\mu_{i}$ are related to the angle defect or surplus
by the relationship $c_{i}=1-4\mu_{i}$ in such manner that if
$c_{i}<1(>1)$ then $\mu_{i}>0(<0)$.
\\
From equation (\ref{curveddirac}) we can write down the equation for
the electron propagator, $S_{F}(x,x')$:

\begin{equation}
i\gamma^{\mu}(x)(\partial_{\mu}-\Gamma^{(T)}_{j\mu})S_{F}(x,x')=
\frac{1}{\sqrt{-g}}\delta^{3}(x-x').\label{diracpropagator}
\end{equation}

The local density of states $N(\omega,\textbf{r})$ is obtained from
the solution of (\ref{diracpropagator}) by Fourier transforming the
time component and taking the limit $\textbf{r}'\rightarrow
\textbf{r}$:
\begin{equation}
N(\omega,\textbf{r})=ImTrS_{F}(\omega,\textbf{r},\textbf{r}).\label{density1}
\end{equation}

Provided that we only consider the presence of pentagons and
heptagons, the parameters $\mu_{i}$ are all equal and small
($\mu_{i}\equiv \mu=1/24$). We will solve equation (\ref{diracpropagator})
perturbatively in $\mu$.
\\
When  dealing with equation (\ref{diracpropagator})
we will reduce the number of spin connections derived in the previous
section by the following considerations: First, we will consider an scenario
where the number of pentagonal and heptagonal defects is the same -
so the total number of defects is even. This suppresses the contribution
from the first exponential in (\ref{composite2}). If we consider that
pentagonal and heptagonal defects come in pairs as usually happens in
the observations, we can neglect the
effect of mixing of the the two sublattices that each individual
odd-sided ring produces and hence eliminate the spin
connection related to $\tau_{2}$ from (\ref{diracpropagator}).
Furthermore, we can disregard the spin connection related to
$\tau_{3}$ by the following argument:
\begin{figure}
  \begin{center}
\includegraphics[height=7cm]{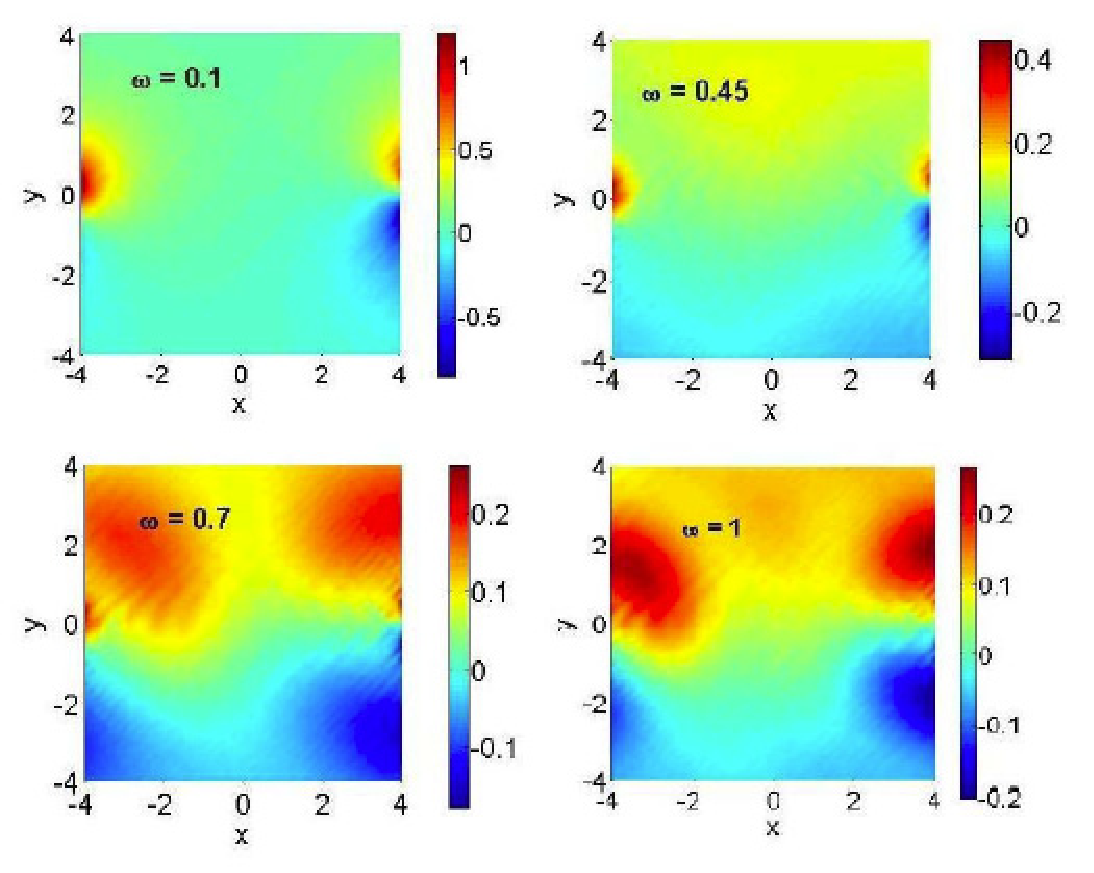}
\end{center}
\caption{First order correction to the local density of states in a
region around two pairs of heptagon-pentagon defects located out of
the image for increasing values of the energy.} \label{densityfig}
\end{figure}
We will solve equation (\ref{diracpropagator}) perturbatively to
first order of the parameter $\mu$. In general if $S^{0}_{F}$ is the
unperturbed Dirac propagator and $\hat{V}(\omega,\textbf{r})$ the
perturbation potential, the first term of such solutions is:

\begin{equation}
S^{1}_{F}(\omega,\textbf{r},\textbf{r}')=\mu \int d^{2}\textbf{r}''
S^{0}_{F}(\omega,\textbf{r},\textbf{r}'')\hat{V}(\omega,\textbf{r}'')
S^{0}_{F}(\omega,\textbf{r}'',\textbf{r}'),\label{perturbsol}
\end{equation}
and we trace $S^{0}_{F}(\omega,\textbf{r},\textbf{r})$ in order to
get the first contribution to the density of states $\delta
N(\omega,\textbf{r})$. The trace operation eliminates all the terms
appearing in (\ref{perturbsol}) which are proportional to a traceless matrix,
including the matrix related to $\tau_{2}$. In fact, up to this
order in perturbation theory, the only term that survives will be
the one proportional to $\gamma^{0}$. With all this in mind, the
relevant spin connection terms are:

\begin{equation}
\Gamma_{1}(\textbf{r})=-\frac{1}{2}\gamma^{1}\gamma^{2}\partial_{y}\Lambda
,
\Gamma_{2}(\textbf{r})=-\frac{1}{2}\gamma^{2}\gamma^{1}\partial_{x}\Lambda.
\label{spinconnection}
\end{equation}

After all  these simplifications
we can write equation (\ref{diracpropagator}) in a
more suitable form.  Expanding the terms in (\ref{spinconnection}) in
powers of $\mu$ we get the potential
$\hat{V}(\omega,\textbf{r})$:

\begin{equation}
\hat{V}(\omega,\textbf{r})=-2\Lambda
\gamma^{0}\omega+i\Lambda\gamma^{j}\partial_{j}+\frac{i}{2}\gamma^{j}(\partial_{j}\Lambda).\label{potential}
\end{equation}
\\
As we said, expression (\ref{perturbsol}) gives us the first
correction to the local density of states in real space. In
figure (\ref{densityfig}) we present an example of the results obtained.
We show the first order correction to the local density of states
coming from two pairs of heptagon-pentagon defects  located out of the image
for increasing values of the energy.
What we see is that as the frequency increases, the local density of states is enhanced
and inhomogeneous oscillations are observed in a wide area around
the defects. The spatial extent of the correction is such that the
relative intensity decays to ten percent in approximately 20 unit
cells. The model described in this work  can be applied to other
configurations of defects, such as simple pairs or stone-Wales
defects. These results can be found in (\cite{CV06a}).
\\
\\
\textit{Acknowledgements}: Funding from MCyT (Spain) through grant
FIS2005-05478-C02-01 and European Union FERROCARBON Contract 12881
(NEST) is acknowledged.

\bibliographystyle{apsrev}
\bibliography{pentagon}

\end{document}